\documentclass[12pt,preprint]{aastex}

\newcommand{\kms}{\mbox{km~s$^{-1}$}}

\shorttitle{Detection of glycolaldehyde in hot molecular cores}
\shortauthors{Beltr\'an et al.}

\begin{document}


\title{First detection of glycolaldehyde outside the
Galactic Center\altaffilmark{1}}

\author{M.\ T.\ Beltr\'an \altaffilmark{2}, C.\ Codella\altaffilmark{3},
S.\ Viti\altaffilmark{4}, R.\ Neri\altaffilmark{5}, R.\ Cesaroni\altaffilmark{6}}


\altaffiltext{1}{Based on observations carried out with
the IRAM Plateau de Bure Interferometer. IRAM is supported by
INSU/CNRS (France), MPG (Germany) and IGN (Spain).}
\altaffiltext{2}{Universitat de Barcelona, Departament d'Astronomia i Meteorologia, Unitat Associada
a CSIC, Mart{\'\i} i Franqu\`es, 08028 Barcelona, Catalunya, Spain;
mbeltran@am.ub.es}
\altaffiltext{3}{INAF, Istituto di Radioastronomia, Sezione di Firenze, Largo
E.\ Fermi 5, I-50125 Firenze, Italy; codella@arcetri.astro.it}
\altaffiltext{4}{Department of Physics and Astronomy, University College London,
Gower Street, London WC1E6BT, UK; sv@star.ucl.ac.uk}
\altaffiltext{5}{IRAM, 300 Rue de la Piscine, F-38406 Saint Martin d'H\`eres,
France; neri@iram.fr}
\altaffiltext{6}{
INAF, Osservatorio Astrofisico di Arcetri, Largo E.\ Fermi 5, I-50125 Firenze,
Italy; cesa@arcetri.astro.it}


\begin{abstract}
Glycolaldehyde is the simplest of the monosaccharide sugars and is
directly linked to the origin of life. We report on the detection of
glycolaldehyde (CH$_2$OHCHO) towards the hot molecular core G31.41+0.31 through IRAM PdBI
observations at 1.4, 2.1, and 2.9~mm. The CH$_2$OHCHO emission 
comes from the hottest ($\ge$ 300~K) and densest ($\ge$2$\times$10$^8$ cm$^{-3}$) region 
closest ($\le$ 10$^4$ AU) to the
(proto)stars. The comparison of data with gas-grain chemical models of hot cores
suggests for G31.41+0.31 an age of a few 10$^5$ yr. We also show that only small
amounts of CO need to be processed on grains in order for existing hot core
gas-grain chemical models to reproduce the observed column densities of
glycolaldehyde, making surface reactions the most feasible route to its
formation.
\end{abstract}



\section{Introduction}

Glycolaldehyde (CH$_2$OHCHO), an isomer of both methyl formate (HCOOCH$_3$) and acetic acid
(CH$_3$COOH), is the simplest of the monosaccharide sugars. The importance of this organic molecule
is that it can react with propenal to form ribose, a central constituent of RNA, and is, therefore,
directly linked to the origin of life. Hence, it is crucial to investigate the occurrence of
glycolaldehyde in the universe, especially in star-forming regions, where stars associated with
planetary systems are expected to form. Interstellar glycolaldehyde has first been detected towards
the Galactic center source Sgr B2(N) by \citet{hollis00}, and successively confirmed
by \citet{halfen06} and \citet{requena08}, 
through single-dish observations. Interferometric observations \citep{hollis01} of
the Sgr B2(N) region, known as the Large Molecule Heimat (LMH), have shown that,
unlike its isomers methyl formate and acetic acid, glycolaldehyde has a large spatial scale,
60$\arcsec$.

\citet{beltran05} have reported the detection of a bright line at 220.46~GHz,
observed with the IRAM Plateau de Bure Interferometer (PdBI), towards the massive star-forming
regions G31.41+0.31 and G24.78+0.08. In the latter case, such a line has been detected in two distinct
cores, A1 and A2. This emission has been interpreted as possibly arising from the transition
20$_{\rm 2,18}$--19$_{\rm 3,17}$ of CH$_2$OHCHO. However, it should be noted that a single tentative
detection of an interstellar molecule transition is not enough for establishing its identification
\citep{snyder05}.
We have performed an observational campaign to detect multiple transitions of CH$_2$OHCHO 
in G31.41+0.31 in order to
confirm for the first time the presence of glycolaldehyde in a compact core in a star
forming region, possibly associated with the material assembling the newly born massive protostars.

G31.41+0.31 is a very massive star-forming region located at a distance of 7.9~kpc
\citep{churchwell90}, associated with a  hot (T $\simeq$ 300 K) molecular core (HMC), with a  
luminosity of $\sim$3$\times10^5 L_\odot$ and thought to be heated by one or more O-B (proto)stars.
\citet{beltran04} have detected a clear velocity gradient perpendicular to the direction of a
bipolar outflow imaged by \citet{olmi96} at the center of the core (see Fig.~\ref{fig2}). Such a
velocity gradient has been interpreted as rotation of a massive toroid ($M_{\rm
gas}$$\sim$490~$M_\odot$) about the axis of the outflow. In addition, recent SMA interferometric
observations at 226~GHz have detected CN and H$_2$CO in red-shifted absorption towards the dust
continuum of the hot core  (Beltr\'an et al., in preparation), which indicates the presence of
accretion in such a massive toroid. Therefore, G31.41+0.31 very much resembles the 20~$M_\odot$
young stellar object G24.78+0.08~A1, where infall, outflow, and rotation have been simultaneously
detected for the first time towards a massive (proto)star \citep{beltran06}.

In this letter, we report the first detection of 
glycolaldehyde towards a HMC located outside the Galactic center, and show that there is a viable
mechanism for glycolaldehyde to be abundant in HMCs by means of chemical modelling.

\section{Observations} 
\label{sobs}

We carried out observations at 1.4, 2.1, and 2.9~mm with the PdBI on March 2003, June 2008,
and July 2008, respectively. Line data have a spectral resolution of 3.4, 0.33, and 0.45~\kms\ at
1.4, 2.1, and 2.9~mm, respectively. Channel maps were created with natural weighting, obtaining
synthesized beams and 1$\sigma$ RMS of 1\farcs1$\times$0\farcs5 and 50~mJy/beam/channel at 1.4~mm,
4\farcs1$\times$3\farcs2 and 13~mJy/beam/channel at 2.1~mm, and 5\farcs4$\times$4\farcs4 and
9~mJy/beam/channel at 2.9~mm.

\section{Detection of glycolaldehyde}

Figure~1 reports the spectra observed at 1.4, 2.1, and 2.9~mm towards the peak position of the HMC
G31.41+0.31. As can be seen in the top panel, besides the CH$_3$CN (12$_8$--11$_8$) and
CH$_3$$^{13}$CN (12$_6$--11$_6$) transitions, other lines are detected.  In particular, there is a
clear emission peak at 220465.86 MHz. Although the line is blended with the CH$_3$CN (12$_{\rm
8}$--11$_{\rm 8}$) and  CH$_3$$^{13}$CN (12$_{\rm 6}$--11$_{\rm 6}$) transitions, there is no doubt
of the presence of a separate line at 220465.86 MHz. On the basis of the Jet Propulsion Laboratory,
Cologne, and Lovas databases for molecular spectroscopy, the only possible identification is the
CH$_2$OHCHO (20$_{\rm 2,18}$--19$_{\rm 3,17}$) line at 220463.87 MHz ($E_{\rm u}$=120 K; $S\mu^2$ =
65.39 D$^2$). The difference between the observed and the laboratory frequencies ($\sim$1.99~MHz),
albeit non negligible, is less than the spectral resolution of 2.50~MHz. Such a discrepancy is not
surprising since it has been found in several  glycolaldehyde transitions observed towards
Sgr~B2(N) \citep{hollis00, halfen06} and could reflect the uncertainties on the CH$_2$OHCHO rest
frequencies in the spectral catalogues. As mentioned above, the glycolaldehyde 20$_{\rm
2,18}$--19$_{\rm 3,17}$ line is blended with  two molecular transitions of CH$_3$CN and
CH$_3$$^{13}$CN. \citet{beltran05}  fitted the three lines with Gaussians having the same line
width (FWHM), and separations in frequency identical to the laboratory values (allowing for the
above mentioned difference of 1.99~MHz for the glycolaldehyde transition). Here, we prefer instead
to leave all parameters free to allow for possible differences in the FWHM of different molecules
and for the sake of consistency with the method adopted to fit the other two glycolaldehyde lines
(see below). The new value of the FWHM of the 1.4~mm line is 9.2$\pm$0.7~km~s$^{-1}$.

\begin{figure}
\begin{center}
\centerline{\includegraphics[angle=0,width=7.6cm]{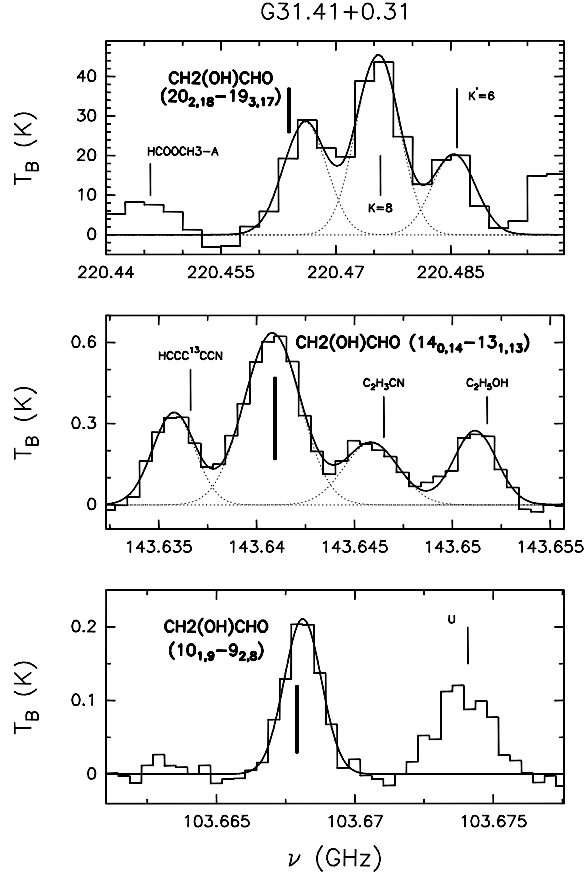}}
\caption{Beam-averaged spectra in $T_{\rm B}$ scale of the CH$_2$OHCHO (20$_{\rm 2,18}$--19$_{\rm 3,17}$),
(14$_{\rm 0,14}$--13$_{\rm 1,13}$), and (10$_{\rm 1,9}$--9$_{\rm 2,8}$)
at 220463.87, 143640.94, and 103667.91 MHz, respectively, as observed towards
the central position of the G31.41+0.31 hot core.
Rest frequencies are pointed out by vertical bars.
{\it Upper panel} (from Beltr\'an et al. 2005): The glycolaldehyde line is blended with
the CH$_3$CN (12--11; $K$=8) line. Two additional lines are present:
(i) $^{13}$CH$_3$CN (12$_{\rm 6}$--11$_{\rm 6}$; labeled by $K^{'}$), and
(ii) HCOOCH$_3$--A (25$_{\rm 11,15}$--26$_{\rm 9,18}$) (2204445.79 MHz;
$E_{\rm u}$=272 K) which could contain an emission contribution due to
the CH$_2$OHCHO (18$_{\rm 4,14}$--17$_{\rm 4,13}$) (220433.51 MHz; $E_{\rm u}$=108 K) line.
The continuous  line shows the fit to the group of three lines formed by
the CH$_2$OHCHO (20$_{\rm 2,18}$--19$_{\rm 3,17}$), CH$_3$CN (12--11; $K$=8)
and $^{13}$CH$_3$CN(12--11; $K^{'}$=6); dotted lines draw the three individual
Gaussian curves used for the fit.
{\it Middle panel}: The CH$_2$OHCHO line is part of a spectral pattern containing also
the HCCC$^{13}$CCN (54--53) (143636.63 MHz; $E_{\rm u}$=183 K),
C$_2$H$_3$CN (33$_{\rm 2,31}$--32$_{\rm 4,28}$ (143646.50 MHz; $E_{\rm u}$=620 K), and
C$_2$H$_5$OH (29$_{\rm 2,28}$--28$_{\rm 3,26}$ (143651.78 MHz; $E_{\rm u}$=415 K) lines.
The results of the fit as drawn as in the Upper panel.
{\it Lower panel}: Besides the glycolaldehyde emission, an unidentified spectral
pattern is present around 103674 MHz. Solid curve shows the fit
of the isolated CH$_2$OHCHO line.}
\label{fig1}
\end{center}
\end{figure}

\begin{figure*}
\begin{center}
\includegraphics[angle=270,width=15cm]{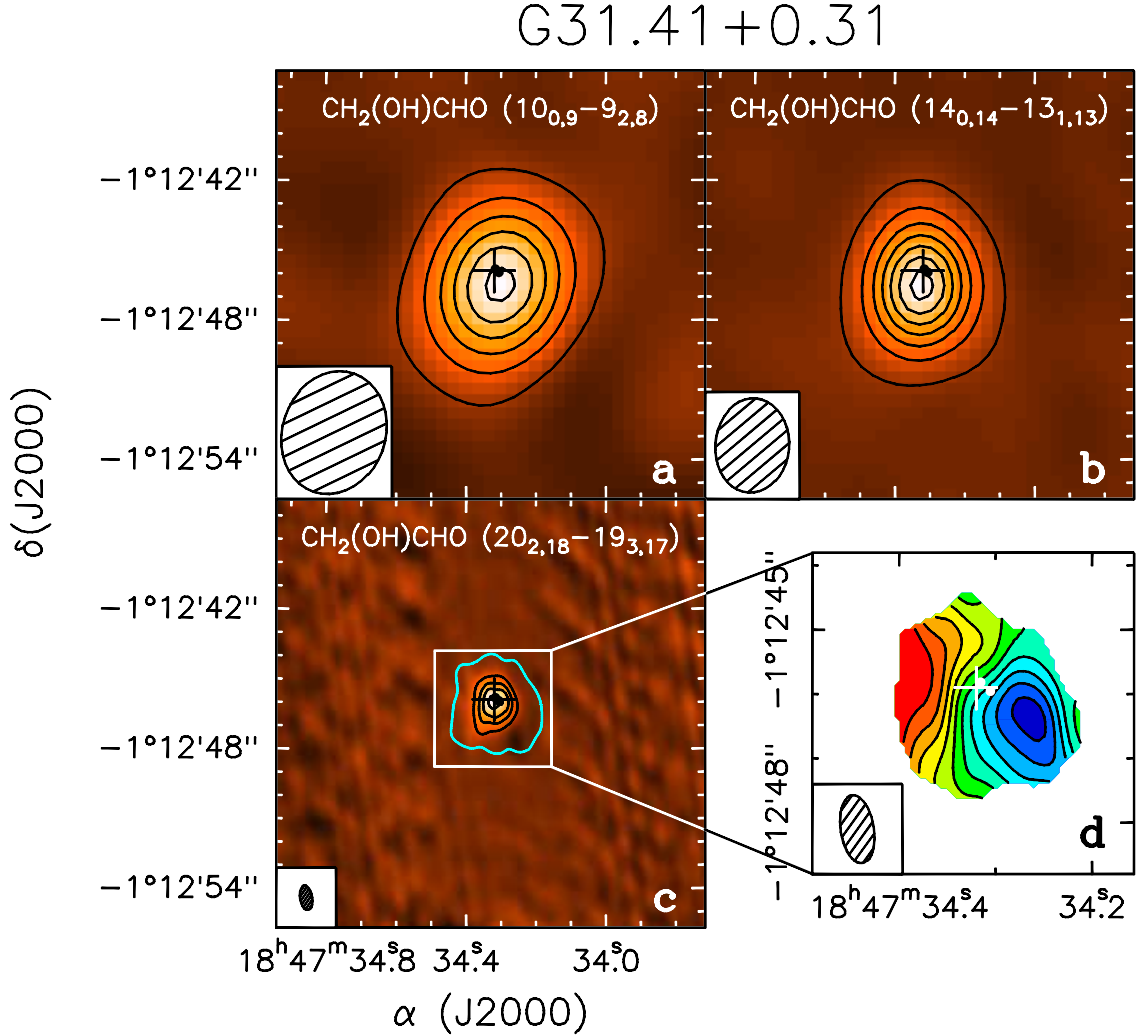}
\caption{Map of the intensity integrated under the  CH$_2$OHCHO
(10$_{0,9}$--9$_{2,8}$) line at 103.67 GHz ({\it a}), the CH$_2$OHCHO
(14$_{0,14}$--13$_{1,13}$) line at 143.64 GHz ({\it b}), and the CH$_2$OHCHO
(20$_{2,18}$--19$_{3,17}$) line at 220.46 GHz ({\it c}), towards the hot
molecular core G31.41+0.31. The 3~$\sigma$ contour of the
CH$_3$CN (12--11) emission averaged under the $K$=0, 1, and 2 components is shown in light blue.
{(\it d)} Close-up of the central region that shows 
the velocity field of the toroid in G31.41+0.31 mapped by \citet{beltran04}. The cross marks the position of the  1.4~mm radio
continuum source, and the dots the 7~mm radio continuum sources \citep{araya08}.  The synthesized beam is shown in the bottom left. The
contour levels are: ({\it a}) from 0.01 to 0.09 in steps of 0.02
Jy\,beam$^{-1}$; ({\it b}) from 0.04 to 0.52 in steps of 0.08 Jy\,beam$^{-1}$;
and ({\it c}) from 0.10 to 0.94 in steps of 0.12 Jy\,beam$^{-1}$.}
\label{fig2}
\end{center}
\end{figure*}

To confirm the tentative detection of glycolaldehyde, we checked for other CH$_2$OHCHO lines that
have already been detected towards the Galactic center. Based on the Galactic center CH$_2$OHCHO
survey performed at 2~mm and 3~mm by \citet{halfen06}, we observed towards G31.41+0.31 the
CH$_2$OHCHO (10$_{\rm 1,9}$--9$_{\rm 2,8}$)  and (14$_{\rm 0,14}$--13$_{\rm 1,13}$) transitions
respectively at 103667.907~MHz ($E_{\rm u}$=32 K; $S\mu^2$ = 26.22 D$^2$) and 143640.936~MHz 
($E_{\rm u}$=53 K; $S\mu^2$ = 67.56 D$^2$). These two transitions are the perfect follow-up for
this search, because their excitation is lower than that of the 220 GHz line and their rest
frequencies are well known. In addition, the CH$_2$OHCHO (14$_{\rm 0,14}$--13$_{\rm 1,13}$) is less
affected by line blending with respect to the 1.4~mm one, whereas the (10$_{\rm 1,9}$--9$_{\rm
2,8}$) line is not contaminated by other spectral emissions. As can be seen in Fig.~\ref{fig1},
both transitions have been clearly detected at the expected frequencies. The derived FWHMs are
4.6$\pm$0.6~km~s$^{-1}$ and 7.1$\pm$0.2 at 103.67 and 143.64~GHz, respectively. Both values are
significantly less than the width of the 1.4~mm line (9.2$\pm$0.7~km~s$^{-1}$). FWHMs spanning of
up to a factor of 2 have also been observed by \cite{halfen06} towards the Galactic center.  In our
case the fact that the line width increases with the energy of the transition could be due to an
excitation effect, if the higher energy lines arise from a smaller, more turbulent region. We can
hence conclude that we have obtained the first detection of glycolaldehyde outside the Galactic
center.

The integrated intensity of the 1.4~mm line, $\int T_{\rm B}$ dv = 201.8$\pm$5.9 K km~s$^{-1}$, is
definitely higher with respect to those of the 2.1~mm (4.8$\pm$0.1 K km~s$^{-1}$) and 2.9~mm
(1.0$\pm$0.1 K km~s$^{-1}$) lines. This is the consequence of the beam dilution effect. Figure~2
shows the maps of the three transitions of CH$_2$OHCHO integrated emission as well as the velocity
gradient detected in CH$_3$CN by \citet{beltran04} towards the center of the hot core. As seen in
the maps, the emission of glycolaldehyde comes from the central region, where the millimeter
continuum peaks.  Our angular resolution at 1.4~mm (1$\farcs$1$\times$0$\farcs$5) is sufficient
to barely resolve  the CH$_2$OHCHO emission (bottom-left panel of Fig.~2), thus minimising beam
dilution effects. On the opposite, the angular resolution of the 2.1 and 2.9~mm maps, $\simeq$
4$''$--5$''$, causes a significant beam dilution, thus reducing the measured brightness
temperature.

Unlike what has been observed towards the Galactic center \citep{hollis01}, the CH$_2$OHCHO emission
does not have a large spatial scale but is rather concentrated towards the center of the core. In fact,
the glycolaldehyde emission has a deconvolved size at the 50\% of the peak of the emission of
$\sim$1$\farcs3$ ($\sim$10300~AU), which is definitely smaller than that derived from standard HMC
tracers, such as e.g. methyl cyanide \citep{beltran05}, which is of $\sim$2$\farcs4$ ($\sim$19000~AU).
As a consequence, as shown by the rotational temperature and column density maps  \cite[see Fig.~7
of][]{beltran05}, CH$_2$OHCHO traces the hottest  (T $\ge$ 300 K) and densest ($n_{\rm H_2}$
$\ge$2$\times$10$^8$ cm$^{-3}$) gas in the surroundings of the newly formed star(s), which is located
closer to the peak of the continuum emission.  

Given the difference of the angular resolution of the three CH$_2$OHCHO maps, the comparison of
the different lines has been done by convolving the 1.4~mm and 2.1~mm observations to the
2.9~mm HPBW (5$\farcs$4$\times$4$\farcs$4). By using the expressions reported by
\citet{hollis00}, we derived the rotational temperature diagram, finding that the diagram is
quite flat, which does not allow to derive the column density from the fit. This is probably
due to the high opacity of the CH$_2$OHCHO lines. In fact, on the one hand, the brightness
temperature of the 1.4~mm line is quite high ($\sim$30~K). On the other hand, after smoothing
all maps to the resolution of the 2.9~mm line, the brightness temperature of the three
transitions turn out to be very similar. This suggests that all three lines are optically
thick. Therefore the beam-averaged column densities obtained from the 2.9 and 2.1~mm lines
assuming T=300~K ($N_{\rm CH_2OHCHO}$ $\simeq$ 3$\times$10$^{15}$~cm$^{-2}$), as well as that
derived from the better resolved 1.4~mm transition ($\sim$10$^{17}$ cm$^{-2}$) should be
considered as lower limits. A rough estimate of the CH$_2$OHCHO abundance has been obtained by
using the H$_2$ column density determined from the continuum dust emission at 1.4~mm
\citep{beltran04}. The estimated abundance is of the order of $10^{-8}$. Such a high value is
consistent with the abundances found by \citet{requena08} towards the Galactic center.      

Besides the detection of glycolaldehyde, we wish to demonstrate not only that its column density is
consistent with the predictions of theoretical models, but also that other species involved in the same
formation route are detected with column densities compatible with those predicted by the models. One of
these is HCOOCH$_3$. It appears that the HCOOCH$_3$ line at 220445.79 MHz cannot be used to
estimate the column density because the value obtained would be implausibly large ($N_{\rm
HCOOCH_3}$$\sim$$10^{20}$~cm$^{-2}$). In fact, according to \citet{turner}, transitions of complex
molecules associated with small ($\le$ 1 D$^2$) $S\mu^2$, often exhibit intensities in excess by orders
of magnitude with respect to the LTE values based on the large $S\mu^2$ lines. We thus decided to estimate the HCOOCH$_3$ column
density using data from \citet{cesa94} for a transition with $S\mu^2$  = 13~D$^2$. The result is
listed in Table~\ref{abun}, after correcting the HCOOCH$_3$ estimates for the different beam with
respect to the CH$_2$OHCHO observations. For this purpose,  the expressions of \citet{requena06}
have been used. Following the same approach, the column density of methanol (CH$_3$OH;
Table~\ref{abun}), another molecule involved in the formation process of glycolaldehyde, has also
been collected from the literature \citep{gibb03}. Note that if HCOOCH$_3$ and CH$_3$OH were
optically thick, the column densities could be much higher. In the following, we compare the column densities thus estimated (see Table~\ref{abun})
with those obtained from theoretical models.

\section{The origin of glycolaldehyde in HMCs}

In this section we briefly investigate whether the column densities of
glycolaldehyde that we derive are indeed
chemically feasible. Our purpose is to investigate whether current
chemical models of HMCs can account for the observed abundance of
glycolaldehyde in massive star forming regions. 

To date, several gas-phase and solid-phase routes of formation for glycolaldehyde have been
proposed \cite[e.g.][]{sorrell01, jalbout07}. A possible route of
formation is via radical reaction of HCO with methanol or with formaldehyde (or, more
likely, via intermediate, e.g.\ CH$_2$OH + HCO, where the first species is a direct product
of methanol). While such reactions would be too slow in the gas phase to produce detectable
abundances of glycolaldehyde in the HMCs lifetime, the high densities (10$^7$ cm$^{-3}$) of
such regions may lead to fast surface reactions: it has been proposed for example \citet{charnley05} that such reactions may occur in close proximity via the hot
secondary electron generated by the passage of a cosmic ray through the ice, or via
photoprocessing of grain mantles by UV starlight which create a high concentration of
radicals in the bulk interior of mantles. Grain-grain collisions then provide excess heat
causing radical-radical reactions to occur and form large organic molecules (Sorrell 2001).
In fact, \citet{bennett07} have performed an experiment where methanol
and CO ices were irradiated with energetic electrons to mimic the presence of cosmic ray
ionizations and found that glycolaldehyde was formed provided that methanol and carbon
monoxide are in close proximity in the ices.

For the purpose of this investigation, we will adopt as route of formation of glycolaldehyde surface
reactions of HCO, H$_2$CO and CH$_3$OH, but we note that these somewhat arbitrary choices do not imply that
other routes are not important. Also note that in this simple model, the formation of glycolaldehyde on ices
via reactions of HCO with methanol implies intermediate passages (just like the formation of methanol from
CO in \citet{viti04} which we assume to be fast. As far as the destruction of glycolaldehyde is
concerned, gas-phase photodissociation would be highly inefficient in such environments due to the high
visual extinctions. Destruction by surface reactions will be taken into account by varying the efficiency of
formation. Glycolaldehyde formation via neutral-neutral reactions on the grains will affect the abundances
of other HMC species such as CH$_3$OH, and HCOOCH$_3$.

We have used a published HMC model \citep{viti04} modified to include the formation reactions for
glycolaldehyde as well as other surface reactions summarized in \citet{bottinelli07} to form complex
surface molecules (Table \ref{tab2}). The model already included hydrogenation on grains.  We have run
a small grid of models where we varied the formation rate coefficients of the reactions forming
CH$_2$OHCHO and related species. Our best fit model is one where the efficiencies of the new reactions
listed in Table \ref{tab2}  need not be large (e.g. only $\sim$ 2\% of CO needs to be converted in
H$_2$CO).  The range of ages indicated by the best fit models is consistent with a typical age
for a HMC ($\sim$10$^5$--5$\times$10$^5$~yr).  In absence of any experimental measurements, we do not know
whether the efficiencies employed for Reactions 2 and 7 (Table~\ref{tab2}) are realistic. 
Nevertheless, we note that the best fitting models imply reasonably small CO conversion
efficiencies and probabilities of formation,  probably feasible on surfaces where grains act
as catalyst. More specifically both reactions occur among products of hydrogenation of CO
hence the necessary mobility is low as all the radicals are already in close vicinity to
each other.

\section{Conclusions}

We report the first detection of glycolaldehyde towards a star forming region, the HMC G31.41+0.31
that is hosting massive young stellar object(s). G31.41+0.31, together with G24.78+0.08 A1, is the
only massive core where evidence of infall, rotation, and outflow have been simultaneously detected.
The maps of CH$_2$OHCHO show that the emission, which is smaller than that derived from standard HMC
tracers, is clearly associated with the most central part of the core. This emission traces the
hottest ($\ge$ 300~K) and densest ($\ge$2$\times$10$^8$ cm$^{-3}$) gas in the surroundings of the
embedded high-mass protostar(s).  Therefore we propose glycolaldehyde as an excellent tracer of   the
innermost regions of HMCs. To detect glycolaldehyde towards other HMCs it is crucial to
carry out high-angular resolution observations because the interferometer will filter out the extended
emission of other lines that might overlap with glycolaldehyde and affect its detectability.

The detection of glycolaldehyde has also profound
implications for the chemistry of star forming regions as it can help to constrain the
evolutionary stage of the core, and therefore of the embedded O-B (proto)stars. In fact, we show
that, for existing HMC gas-grain chemical models, the age of the HCM has to be
a few 10$^5$~yr to reproduce the observed column densities of
glycolaldehyde (as well as methyl formate and methanol). 
In addition, only small amounts of CO ($\sim$10--15\%) need to be processed on grains, making surface reactions the most feasible route to
the formation of glycolaldehyde.

\acknowledgments
We thank Prof D. A. Williams for helpful discussion as well as a thorough reading 
of the manuscript. SV acknowledges financial support from an individual PPARC Advanced Fellowship.







\clearpage

\begin{table}
\caption[] {Source-averaged column densities towards the G31.41+0.31 HMC.}
\label{abun}
\begin{tabular}{ccc}
\hline
\multicolumn{1}{c}{CH$_2$OHCHO\tablenotemark{a}}
&\multicolumn{1}{c}{{HCOOCH$_3$\tablenotemark{a}}}&
\multicolumn{1}{c}{CH$_3$OH\tablenotemark{b}} 
\\ 
\multicolumn{1}{c}{(cm$^{-2}$)}
&\multicolumn{1}{c}{(cm$^{-2}$)}
&\multicolumn{1}{c}{(cm$^{-2}$)} 
\\
\hline
$>$1$\times$10$^{17}$\tablenotemark{c} & 3.4$\times$10$^{18}$\tablenotemark{d,\, \rm e} &
7.5$\times$10$^{17}$\tablenotemark{e} \\
\end{tabular}
\tablenotetext{a}{Assuming a kinetic temperature of 300 K, as measured by \citet{beltran05}.}
\tablenotetext{b}{From \citet{gibb03}.}
\tablenotetext{c}{From the CH$_2$OHCHO (20$_{2,18}$--19$_{3,17}$) data.}
\tablenotetext{d}{From the HCOOCH$_3$--E (9$_{6,3}$--8$_{6,2}$) (110652.89 MHz; $E_{\rm u}$=51 K) data of \citet{cesa94}.}
\tablenotetext{e}{In case of optically thick emission, the column densities could be higher.}
\end{table}

\clearpage

\begin{table}
\caption{List of reactions added to the Viti et al.~(2004) model.
	 ``M'' denotes species in the solid phase.}
\begin{tabular}{|c|c|}
\hline
$N$ & Reaction \\
\hline
1 & CO + 4(MH)  $\Rightarrow$ MCH$_3$OH \\ 
2 & CO + MCH$_3$OH $\Rightarrow$ MHCOOCH$_3$ \\
3 & H$_2$CO + MH $\Rightarrow$ MCH$_3$O \\
4 & MCH$_3$O + MHCO $\Rightarrow$ MHCOOCH$_3$ \\
5 & CO + 2(MH)$\Rightarrow$ MH$_2$CO \\
6 & CO + MH $\Rightarrow$ MHCO \\
7 & MH$_2$CO + MHCO + MH $\Rightarrow$ MCH$_2$OHCHO \\
\hline
\end{tabular}
\label{tab2}
\end{table}

\clearpage

\clearpage

\begin{thebibliography}{}

\bibitem[Araya et al.(2008)]{araya08} Araya, E., Hofner, P., Kurtz, S., Olmi, L., \& Linz, H.\
2008, ApJ, 675, 420

\bibitem[Beltr\' an et al.(2006)]{beltran06} Beltr\' an, M.\ T., Cesaroni, R., Codella, C., Testi, L., Furuya,
R.\ S., \& Olmi, O.\ 2006, Nature, 443, 427 

\bibitem[Beltr\' an et al.(2004)]{beltran04} Beltr\' an, M.\ T., Cesaroni, R., Neri, R., Codella, C., Furuya,
R.\ S., Testi, L., \& Olmi, O.\ 2004, ApJ, 601, L190 

\bibitem[Beltr\' an et al.(2005)]{beltran05} Beltr\' an, M.\ T., Cesaroni, R., Neri, R., Codella, C., Furuya,
R.\ S., Testi, L., \& Olmi, O.\ 2005, A\&A, 435, 901 

\bibitem[Bennet \& Kaiser(2007)]{bennett07}
 Bennett, C.\ J., \& Kaiser, R.\ I.\ 2007, ApJ, 661, 899 

\bibitem[Bottinelli et al.(2007)]{bottinelli07} 
Bottinelli, S., Ceccarelli, C., Williams, J.\ P., \& Lefloch, B.\ 2007, A\&A, 463, 601

\bibitem[Cesaroni et al.(1994)]{cesa94}
Cesaroni, R., Olmi, L., Walmsley, C.\ M., Churchwell, E., \& Hofner, P.\ 1994, ApJ, 435, L137

\bibitem[Charnley \& Rodgers(2005)]{charnley05}
Charnley, S.\ B., \& Rodgers, S.\ D.\ 2005, in Astrochemistry: Recent Successes and Current Challenges, 
IAU Symposium S231, eds.\ D.\ C.\ Lis, , G.\ A.\
Blake, \& E.\ Herbst, (Cambridge University Press, Cambridge), 237

\bibitem[Churchwell, Walmsley, \& Cesaroni(1990)]{churchwell90}
Churchwell, E., Walmsley, C.\ M., \& Cesaroni, R.\ 1990, A\&AS, 83, 119

\bibitem[Gibb, Wyrowski, \& Mundy(2003)]{gibb03} Gibb, A.\ G., Wyrowski, F., \& Mundy, L.\ G.\ 2003, in SFChem
2002: Chemistry as a Diagnostic of Star Formation, eds.\ C.\ L.\ Curry \& M.\ Fich, (NRC Press, Ottawa), 214

\bibitem[Halfen et al.(2006)]{halfen06} Halfen, D.\ T., Apponi, A.\ J., Woolf, N., Polt, R., \&
Ziurys, L.\ M.\ 2006, ApJ, 639, 237

\bibitem[Hollis et al.(2000)]{hollis00} Hollis, J.\ M., Lovas, F.\ J., \& Jewell, P.\ R.\ 2000, ApJ,
540, L107

\bibitem[Hollis et al.(2001)]{hollis01} Hollis, J.\ M., Vogel, S.\ N., Snyder, L.\ E., Jewell, P.\
R., \& Lovas, F.\ J.\ 2001, ApJ, 554, L81

\bibitem[Jalbout(2007)]{jalbout07}
Jalbout, A.\ F.\ 2007, Molecular Physics, 105, 941 

\bibitem[Olmi et al.(1996)]{olmi96} Olmi, L., Cesaroni, R., \& Walmsley, C.\ M.\ 1996, A\&A, 307, 599

\bibitem[Requena-Torres et al.(2008)]{requena08} Requena-Torres, M.\ A., Mart\'{\i}n-Pintado, J.,
Mart\'{\i}n, S., \& Morris, M.\ R.\ 2008, ApJ, 672, 352

\bibitem[Requena-Torres et al.(2006)]{requena06} Requena-Torres, M.\ A., Mart\'{\i}n-Pintado, J.,
Rodr\'{\i}guez-Franco, A., Mart\'{\i}n, S., Rodr\'{\i}guez-Fern\'andez N.\ J., \& de Vicente, P. 
2007, ApJ, 971, 455 

\bibitem[Snyder et al.(2005)]{snyder05}
Snyder, L.\ E., Lovas, F.\ J., Hollis, J.\ M., Friedel, D.\ N.\ et al.\ 2005, ApJ, 619, 914  

\bibitem[Sorrell(2001)]{sorrell01}
Sorrell, W.\ H.\ 2001, ApJ, 555, L129

\bibitem[Turner(1991)]{turner}
Turner, B.\ E. 1991, ApJS, 76, 617

\bibitem[Viti et al.(2004)]{viti04}
Viti, S., Collings, M.\ P., Dever, J.\ W. McCoustra, M.\ R.\ S., \& Williams, D.\ A.\ 2004, MNRAS,
354, 1141

\end{thebibliography}
\end{document}